\documentclass[twocolumn]{aastex631}
\usepackage{subfigure}

\begin{document}

\title{The Optical Phase Curves of CoRoT-1 b}

\author{Andrew Li}
\affiliation{No Academic Affiliation\footnote{The author is currently a high school student. \newline contact email - andrewli6907@gmail.com}}

\begin{abstract}

Of the three space telescopes launched so far to survey transiting extrasolar planets, CoRoT is unique in that it was the only one with spectral resolution, allowing for an extraordinary opportunity to study the reflective properties of exoplanets at different wavelengths. In this work, I present a systematic lightcurve analysis of the white-light and chromatic CoRoT lightcurves of CoRoT-1 in order to search for the secondary eclipse and orbital phase variation of the transiting extrasolar planet CoRoT-1 b, as well at search for any chromatic difference in the aforementioned effects. I manage to detect a significant secondary eclipse in the white lightcurve, and detect the eclipse marginally in all three of the color channels. However I am only able to significantly detect the planetary phase variation in the red channel lightcurve. The retrieved secondary eclipse depth is higher in the blue and green channels compared to the white and red, suggesting that CoRoT-1 b has a higher geometric albedo at shorter wavelengths. I also attempt to detect the secondary eclipse using TESS, but show that the available volume and precision of the data is not high enough to allow detection of the secondary eclipse.

\end{abstract}

\section{Introduction} \label{sec:intro}

Since the commencement of observations in February 2006, the Convection, Rotation, and planetary Transits satellite (CoRoT) continuously monitored several fields of the night sky for six years in order to search for transiting extrasolar planets, becoming the first space telescope of its kind launched for such a purpose. One of the advantages of such continuous monitoring of transiting exoplanet systems is that it allows for high precision photometry using relatively small optical components, with the high optical precision being achieved by the large volume and time baseline of the data. As a result, the study of the optical phase curves of transiting hot Jupiter-type exoplanetary systems, which require a high photometric precision, has benefited enormously from the advent of CoRoT, and its successor survey telescopes, Kepler and TESS. Specifically, observations of the planetary secondary eclipse and phase curve are only possible with the photometric precision afforded by space telescopes.

Phase curves of exoplanets usually are studied at infrared wavelengths, both because the phase curve has a higher amplitude at infrared wavelengths, and the fact that observing the infrared phase curves of exoplanets allows for the modelling of the planet's atmospheric structure and composition. However, at optical wavelengths, the phase curve instead provides information about the reflective properties of the exoplanet, including the geometric albedo of the planet and the homogeneity or existence of clouds in the planet atmosphere (\citet{2015AJ....150..112S}).

However, one measurement at one wavelength is usually not enough to model or constrain the reflective and thermal properties of an exoplanet. Because of this, numerous telescopes from both space and the ground have been used to study the secondary eclipses and phase curves of hot Jupiters in the infrared, allowing for the detailed modelling of exoplanet thermal emission spectra.

The same cannot be said for visible-light measurements, where detailed multi-wavelength secondary eclipse observations of hot Jupiters in the optical are few and far between, especially at wavelengths shorter than 500 nm. At the moment, detailed secondary eclipse observations of exoplanets shortward of 500 nm have only been conducted for nine\footnote{The author is aware of four additional exoplanets with unpublished secondary eclipse observations shortward of 500 nm, including a previous analysis of CoRoT-1 b.} exoplanets, with only one of these attempts resulting in a successful detection.
\begin{deluxetable}{ccc}
\tablecaption{Previous Secondary Eclipse Measurements with $\lambda < 500 nm$}
\tablenum{1}
\tablehead{\colhead{Planet} & \colhead{$\delta_{occ}$ (ppm)} & \colhead{Reference} \\ 
\colhead{} & \colhead{} & \colhead{} } 
\startdata
CoRoT-2 b & - & \citet{2010AA...513A..76S} \\
HD 189733 b & $126_{-37}^{+36}$ & \citet{2013ApJ...772L..16E} \\
KELT-9 b & $-71 \pm84$ & \citet{2018ApJ...869L..25H} \\
TrES-3 b & $80 \pm90$ & \citet{2022RNAAS...6..182M} \\
WASP-12 b & $59 \pm134$ & \citet{2019AJ....158...91S} \\
WASP-19 b & $10 \pm280$ & \citet{2012ApJS..201...36B} \\
WASP-43 b & $<860 (3\sigma)$ & \citet{2014AA...563A..40C} \\
 & $-70 \pm110$ & \citet{2022RNAAS...6..182M} \\
WASP-46 b & $490 \pm300$ & \citet{2014AA...567A...8C} \\
WASP-103 b & $200 \pm160$ & \citet{2022RNAAS...6..182M} \\
\enddata
\tablecomments{\citet{2010AA...513A..76S} did not detect the secondary eclipse of CoRoT-2 b in the blue CoRoT channel, but did not provide an upper limit value.}
\end{deluxetable}

CoRoT-1 b is an exoplanet in a unique position ripe for chromatic analysis of its secondary eclipses and planetary phase curves. It possesses an orbital configuration that makes the expected eclipse depth relatively large, and the continuous white-light and chromatic monitoring of the host star by CoRoT means that there is ample data to make a detection of the eclipse and phase curve. Indeed, previous attempts to detect the secondary eclipse have proven that the eclipse depth and phase curve is large and easily detectable with the CoRoT data (\citet{2009A&A...506..353A}, \citet{2009Natur.459..543S}). Taking all of the factors that make this planet conducive to study of its eclipse and phase curve, I endeavor to try and detect these effects in the white-light and chromatic CoRoT lightcurves.

In addition to the CoRoT lightcurves of CoRoT-1, CoRoT-1 and its transiting planet have also been observed by the Transiting Exoplanet Survey Satellite (TESS). In an effort to add another data point to the chromatic eclipse depths of CoRoT-1 b, I also attempt to detect the eclipse in TESS data.

\section{Methods} \label{sec:methods}
I use the public N2 CoRoT data for CoRoT-1 in .tbl format, as provided by the NASA Exoplanet Archive. The white-light and chromatic lightcurves are then extracted from the lightcurve file. To allow a uniform analysis of all of the lightcurve, the section with 32s candence was split off from the previous section with 512s cadence, and binned into bins of 16 points to match the rest of the data. The resulting lightcurve consists of 7117 datapoints spread across 54.72 days and 36.26 planetary orbits.

The first and most obvious systematic effect in the CoRoT lightcurves are the numerous flux discontinuities, or "jumps", which are caused by the impact of high-energy particles onto the spacecraft's CCD detectors. I designed a fairly basic but effective way of removing these discontinuities from the lightcurve by taking the average flux of the lightcurve 0.5 days before and after the flux jump and subtracting the differences in the means from the part of the lightcurve that occurs after the jump. The resulting lightcurves were much smoother than the originals.
\begin{deluxetable}{ccc}
\tablecaption{Adopted Planetary Parameters}
\tablenum{2}
\tablehead{\colhead{Parameter} & \colhead{Value} & \colhead{Reference} \\ 
\colhead{} & \colhead{} & \colhead{} } 
\startdata
$a/R_*$ & 4.751 & \citet{2019AA...628A.115V} \\
$R_P/R_*$ & 0.1419 & \citet{2019AA...628A.115V} \\
$P$ & 1.508968772 d & \citet{2022ApJS..259...62I} \\
$T_*$ & 6355 K & \citet{2019AJ....158..138S} \\
$T_{P, day}$ & 2279 K & \citet{2023AJ....165..104D} \\
\enddata
\tablecomments{The cited $T_{P, day}$ value is the average of the brightness temperatures in the 3.6 and 4.5 micron bands.}
\end{deluxetable}

For the TESS lightcurves, I use the SPOC 120s cadence Sector 6 and 33 lightcurves as provided by the download function of the Python package lightkurve (\citet{2018ascl.soft12013L}). I extracted the PDC lightcurve from these files, as compared to the SAP lightcurve, the instrumental effects impacting the SAP lightcurve have been removed in the PDC lightcurve. The lightcurves from each of the two sectors were then joined and analyzed together. Because the time gap between the two sectors' data is large, I do not expect this action to significantly affect the detrending process or the end results of the analysis. The resulting lightcurve contains 30298 data points, is near continuous over 47.61 days, and covers 31.55 planetary orbits, with a large gap in time between the Sector 6 and Sector 33 data.

The TESS and CoRoT lightcurves were then detrended using the Python package wotan (\citet{Hippke_2019}). In the case of the CoRoT data, I used a biweight filter with a window size equal to that of the planetary orbital period. This is chosen specifically because, in this way, the detrending algorithm will detrend out all long term trends across longer timescales than the planetary orbital period while keeping any possible planetary phase variation in the detrended lightcurve. In the case of the TESS data, I initially also processed the TESS data in the same manner, but discovered that, because of the faintness of the host star, correlated noise dominated the resulting lightcurve and its analysis. As a result, I decided to utilize a smaller window of about 0.7 days. This was too small to detrend over the planetary transits properly, so the transits in the TESS lightcurves were removed before the detrending process was applied. This means that the phase curve will be detrended out of the lightcurve, and that I will no longer be able to search for it in the TESS data. However, the secondary eclipse should still be present, since it is a high frequency feature on a scale much smaller than the detrending window.

The resulting TESS and CoRoT lightcurves were then $\sigma$-clipped to $3\sigma$, which had the effect of clipping out not only any possible outliers, but also the planetary transits in the CoRoT lightcurves as well. After detrending was performed, it became obvious that several low-level flux modulations and residual incompletely removed discontinuities remained. I tested a variety of methods to remove these effects, but, in accordance with the methods of previous analyses such as \citet{2009Natur.459..543S}, I found that the best way to account for these effects was to just remove them from the lightcurve altogether. To do this, I constructed a "blacklist" of ranges of data points, determined by a visual inspection of the data, and then automatically removed these sections from the sigma-clipped data. In the end, the red and green channel CoRoT lightcurves, as well as the TESS lightcurve, did not have a blacklist, meaning that I did not find any portions of those respective lightcurves that warranted removal.

\begin{deluxetable}{ccc}[htb!]
\tablecaption{CoRoT-1 b Lightcurve Blacklist}
\tablenum{3}
\tablehead{\colhead{Color} & \colhead{Interval} & \colhead{Reason for Removal} \\ 
\colhead{} & \colhead{} & \colhead{} } 
\startdata
WHITE & \{2594.0, 2596.2\} & flux jump \\
WHITE & \{2598.5, 2599.0\} & flux jump \\
WHITE & \{2622.0, 2623.0\} & undetrended flux variation \\
WHITE & \{2635.5, 2637.7\} & flux jump \\
WHITE & \{2640.0, 2640.8\} & flux jump \\
WHITE & \{2642.5, 2644.4\} & flux jump \\
BLUE & \{2594.0, 2595.0\} & flux jump \\
BLUE & \{2598.5, 2599.0\} & undetrended flux variation \\
BLUE & \{2637.0, 2637.7\} & flux jump \\
BLUE & \{2642.5, 2644.4\} & flux jump \\
\enddata
\end{deluxetable}
After the previously mentioned processing steps, the TESS and CoRoT white lightcurves are ready to be analyzed for the presence of planetary phase variations and secondary eclipses. However, the CoRoT color lightcurves need additional processing. This is because the color channel lightcurves, unlike the white lightcurve, are not corrected for systematics caused by pointing drifts of the telescope. Two main systematic effects are visible in the color channel lightcurves: a 103-min high-amplitude variation cognate with the satellite orbital period, and a 24-hour low-amplitude sinusoidal variation. I remove these variations in largely the same way that is done by \citet{2009Natur.459..543S}. A sinusoidal function with a period of 24 hours was fitted to the data, with the resulting lightcurve being the residuals of the fit. Then, the data was then phase-folded to the 103-min orbital period of the satellite, and then boxcar-smoothed with a window of 8 datapoints. The resulting curve was then used to detrend the entire lightcurve so that no systematic variation remained. I should note that \citet{2009Natur.459..543S} find evidence of a \(\sim \)10 day variation in the data, which they attribute to spot variability and/or stellar rotation. I do not attempt to find and remove this variability, because such a large-period variation is likely to have already been removed by the detrending algorithm, and the estimated effect of such a low-amplitude variation on the final values is ostensibly small.

After all detrending was accomplished, the lightcurves were then phase-folded to the planetary orbital period. I then tested fitting two different models to the CoRoT lightcurves - one a pure phase curve and eclipse model, and the other including the eclipse and all three phase curve effects, reflection and thermal emission, ellipsoidal variation, and Doppler beaming. Note that I do not expect to detect the ellipsoidal variation and Doppler beaming in the lightcurves, since the expected amplitude is quite small compared to the photometric precision of the lightcurves, but instead use the ellipsoidal variation and doppler beaming amplitudes as "correlated noise detrending functions". The rationale behind this is that, due to the nature of the color channel lightcurves, it is likely that significant correlated noise remains in the lightcurves. The "ellipsoidal variation" and "Doppler beaming" curves would therefore be able to eliminate the two sources of noise most likely to influence the phase curve amplitude: noise at phases 0.25 and 0.75, and noise causing the first half of the phase curve to be higher than the second half. After the fitting was complete, a visual inspection of the fit was done, and based on this visual inspection I adopted the "reflection-only" model for the white lightcurve and the red and green color lightcurves. For the blue-channel lightcurve, the "3-effect" model provided a better fit, so I adopted this model for the blue channel lightcurve instead. In the case of the TESS lightcurve, where I do not expect any contribution from a putative phase curve signal due to the nature of the lightcurve detrending, I simply adopted an eclipse-only fit.

The fitting was done using a DLS (damped least-squares) method, and the error estimation was done using a residual-shuffling bootstrap method similar to the one used in \citet{2009A&A...506..353A}. I should note that such a method does not account for additional uncertainty due to correlated noise, and as a result I expect that my derived error bars are marginally underestimated. Ideally, analysis of this data should be done using an MCMC (Markov-chain Monte Carlo) parameter and uncertainty estimation method, which is more robust to different sources of uncertainty than the types of bootstrap methods described here, but unfortunately, due to the limited computational resources of the author, I am unable to perform an MCMC analysis at this time. Despite the limitations of my analysis, a visual inspection of the fitted model suggest that the analysis is robust.

\section{Results} \label{sec:results}
From my analysis, I clearly detect the secondary eclipse in the white (4.44$\sigma$) channel data, and marginally detect the eclipse in the three color channels red (2.72$\sigma$), green (2.99$\sigma$), and blue (2.44$\sigma$). The eclipse is not detected in the TESS data, and I am only able to set a $1\sigma$ upper limit of 223 ppm for a possible eclipse depth. As for the planetary phase variation, I clearly detect the planetary phase variation only in the red channel (3.01$\sigma$), with only marginal detections in the white (2.50$\sigma$), green (1.23$\sigma$), and blue (2.41$\sigma$) channels. As mentioned before, I do not expect to detect the planetary phase variation in the TESS lightcurve due to the nature of the detrending algorithm. According to \citet{parviainen_deeg_belmonte_2013}, the flux contamination in the CoRoT lightcurves is negligible, and the TESS PDC lightcurves come pre-adjusted for contamination, so no further correction was required.

To verify that the detected eclipse signal is real, I shifted the eclipse model in units 0.01 in phase between phases 0.25 and 0.75, and fitted each model with its own corresponding eclipse depth. The resulting array of values show a spike in the eclipse depth at phase 0.5 and a lack of periodic variability elsewhere, suggesting that my detection of the eclipse depth is real and not due to correlated noise or undetrended systematics.
\begin{deluxetable}{ccc}[htb!]
\tablecaption{CoRoT-1 b Phase Curve Fits}
\tablenum{4}
\tablehead{\colhead{Color} & \colhead{$\delta_{occ}$ (ppm)} & \colhead{$A_{Refl}$ (ppm)} \\ 
\colhead{} & \colhead{} & \colhead{} } 
\startdata
WHITE & 192 ±43 & 56 ±22 \\
RED & 147 ±54 & 82 ±27 \\
GREEN & 334 ±112 & 58 ±47 \\
BLUE & 323 ±132 & 110 ±46 \\
TESS & $<223$ ($1\sigma$) & \\
\enddata
\end{deluxetable}
\begin{figure*}[hbt!]
\plotone{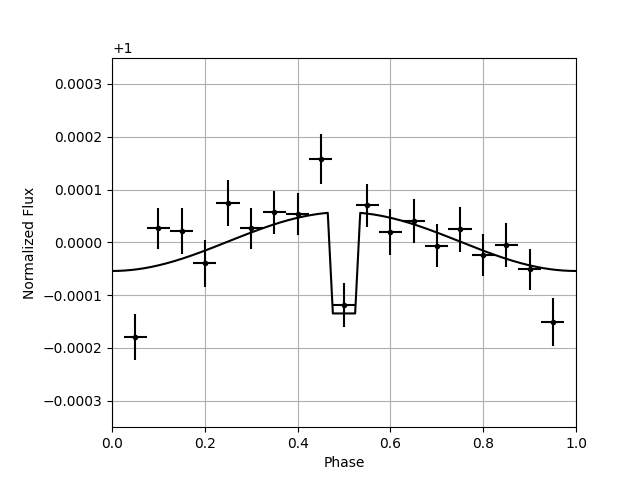}
\caption{CoRoT white-light phase curve of CoRoT-1 b. 
\label{fig:white}}
\end{figure*}
\begin{figure*}[hbt!]
\plotone{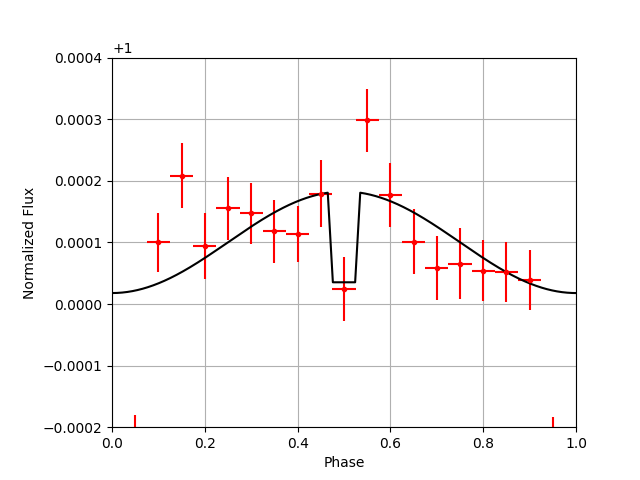}
\caption{CoRoT red channel phase curve of CoRoT-1 b. 
\label{fig:red}}
\end{figure*}
\begin{figure}[hbt!]
\plotone{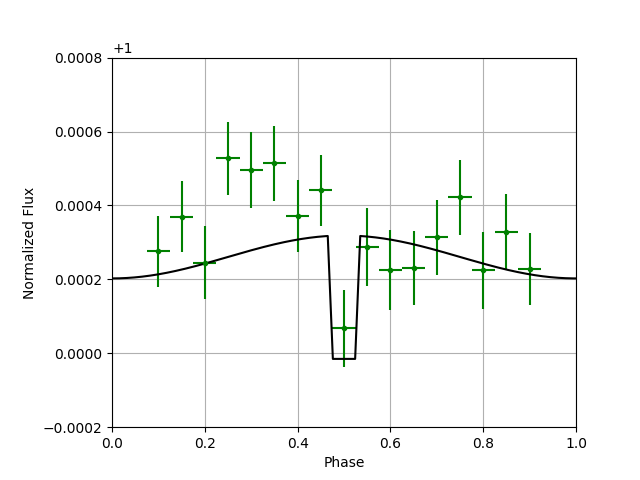}
\caption{CoRoT green channel phase curve of CoRoT-1 b. 
\label{fig:green}}
\end{figure}
\begin{figure}[hbt!]
\plotone{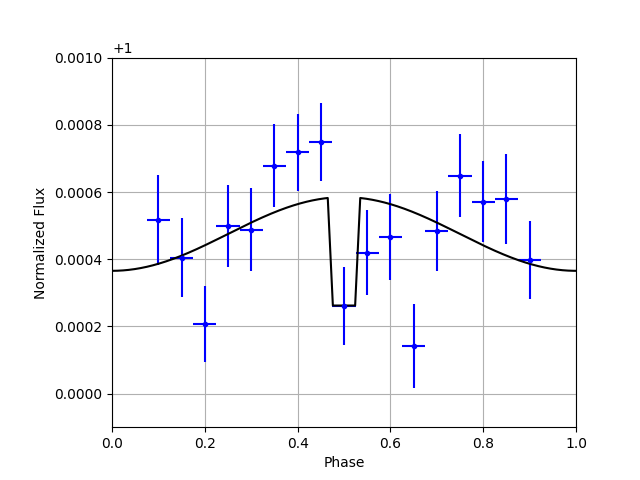}
\caption{CoRoT blue channel phase curve of CoRoT-1 b. 
\label{fig:blue}}
\end{figure}
\begin{figure}[hbt!]
\plotone{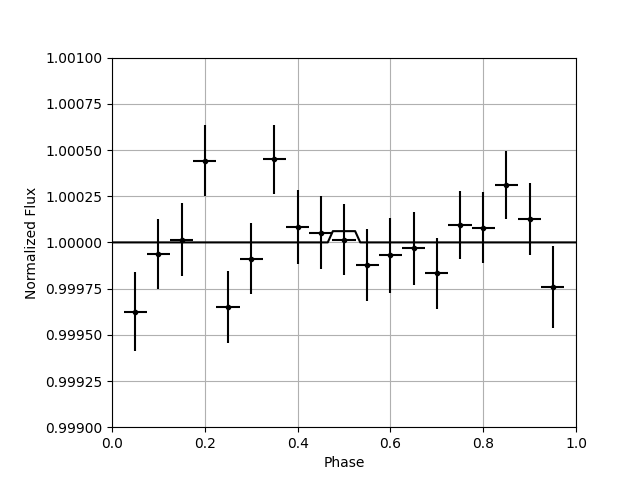}
\caption{TESS white-light phase curve of CoRoT-1 b. 
\label{fig:tess}}
\end{figure}

\section{Discussion} \label{sec:discussion}
\subsection{Significance of Results}
For each of the four CoRoT color channels, I measure the significance of the differences between the phase curve amplitude and the secondary eclipse depth (or in other terms, the nightside flux) to be 1.30$\sigma$ (white), 0.23$\sigma$ (red), 1.49$\sigma$ (green), and 0.64$\sigma$ (blue).
The apparently significant detection of the nightside flux in the white and green channels is probably not indicative of real nightside emission, as such emission would not be detectable at the wavelengths of the white and the green channels. Instead, the difference probably indicates that the apparent phase curve signal has been partially masked by unmodeled systematics and other effects caused by pointing drifts of the CoRoT spacecraft, and that these values of the phase curve amplitudes should not be taken at their face value. Thus, I can only confidently conclude that I have detected the phase variation in the red channel, and marginally in the blue channel, while in the white lightcurve I probably have also detected the phase variation of the planet, but that I cannot determine the true value of the phase amplitude due to residual systematic effects. In the case of the green channel, any possible phase variation has been completely wiped out by the noise in the data.

My derived white-light eclipse depth is consistent with the value found by \citet{2009A&A...506..353A}, while my red channel eclipse depth is marginally larger than the value found by \citet{2009Natur.459..543S}, but is still somewhat consistent. I attribute these differences to the difference in data processing.
\subsection{Solving the CoRoT Color Channels}
Calculations of the wavelength ranges of the channels is required to place these measurements into their proper context. The CoRoT color channels do not correspond to any conventional photometric system, and are actually different for each star. This is because the color channels are formed by a small bi-prism that is placed in front of the Planet-Finder channel, which disperses the light slightly. A photometric mask is then applied to the dispersed flux, and based on the relative intensities of the photons and the position of the star on the CCD, the three color channels are created. While the exact ratio of red to green to blue flux is nontrivial to calculate, \citet{RabelloSoares_2022} use empirically calculated values to approximate the "average" wavelength properties of the color channels, which I use for the purposes of this analysis. To calculate the wavelength limits of the color channels, I multiply a calculated Planck curve of the star by the CoRoT CCD response function from \citet{auvergne_2009}, then took the wavelength cutoff with the bluest 22\% of the flux to be the blue channel cutoff, and the wavelength cutoff with the reddest 63\% of the flux to be the red channel cutoff. From these calculations, I calculate the blue wavelength cutoff to be 502 nm, and the red wavelength cutoff to be 561 nm, and the corresponding effective wavelengths of the white, blue, green, and red channels to be 623 nm, 456 nm, 533 nm, and 703 nm. This is in agreement with \citet{2009Natur.459..543S}, who find an effective wavelength of the red channel for CoRoT-1 of 710 nm and a cutoff of 560 nm.

It would also be appropriate to conduct an analysis of the effective wavelength of the TESS lightcurve, but given that I do not detect the secondary eclipse in the TESS data, and the fact that most other publications have not adjusted the TESS bandpass to account for the stellar spectrum, I decide not to proceed for consistency with other literature analyses of TESS eclipses. In any case, the potential impact on the geometric albedo upper limit due to this decision is expected to be small due to the relatively wide bandpass of the TESS cameras.

\begin{figure}[bt!]
\plotone{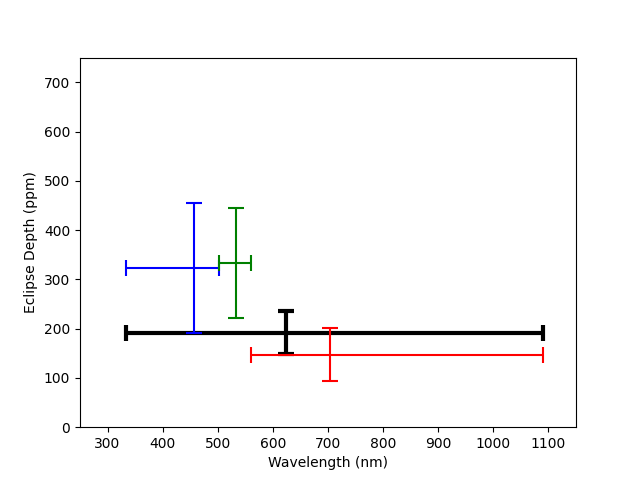}
\caption{CoRoT white-light and chromatic secondary eclipse depths. 
\label{fig:general}}
\end{figure}
\subsection{A Wavelength-Dependent Albedo of CoRoT-1 b?}
Provided that the dayside temperature of the planet, determined from infrared observations of the secondary eclipse, is known, one can calculate the geometric albedo (corrected for the presence of thermally emitted light) of the exoplanet under the assumption that the star and the planetary dayside radiate like black bodies using the following equation:
\begin{equation}
A_g = \delta_{occ} (\frac{a}{R_P})^2 - \frac{B (\lambda, T_{P, day})}{B (\lambda, T_*)} (\frac{a}{R_*})^2
\end{equation}

I take the derived dayside brightness temperatures from \citet{2023AJ....165..104D} and average the 3.6 micron and 4.5 micron temperatures to arrive at an approximate estimate of the planetary dayside temperature to use in calculating the albedo. Thus, the thermally corrected albedos of CoRoT-1 b can be calculated.
\begin{deluxetable}{ccc}[htb!]
\tablecaption{Geometric Albedos}
\tablenum{5}
\tablehead{\colhead{Color} & \colhead{$A_g$} & \colhead{$3\sigma$ 
 upper limit}\\ 
\colhead{} & \colhead{}} 
\startdata
WHITE & 0.182 ±0.048 & - \\
RED & 0.097 ±0.061 & $<0.280$ \\
GREEN & 0.363 ±0.125 & $<0.738$ \\
BLUE & 0.359 ±0.148 & $<0.803$ \\
TESS & $<0.155$ ($1\sigma$) & $<0.655$ \\
\enddata
\end{deluxetable}

The geometric albedo of CoRoT-1 b appears to be elevated in the green and blue channels, in stark contrast to most other hot Jupiter-type planets, which typically have very low albedos (\citet{Esteves_2015}). This suggests the presence of reflective clouds on the dayside of CoRoT-1 b. In addition to the elevated geometric albedo, the albedo in the green and blue channels is higher than the albedo in the red channel, as well as the overall white channel albedo (which predictably is in between the albedos of the red channel and the green and blue channels), and the $1\sigma$ upper limit of the albedo in the TESS bandpass. This suggests some chromatic dependence in the geometric albedo of CoRoT-1 b, possibly due to Rayleigh scattering, or the inclusion of pigmenting and/or scattering compounds in the clouds. This chromatic difference in the albedo means that, to the human eye, CoRoT-1 b would likely appear to be bluish in color. However, I should caution that the significance of the eclipse depths and albedos I derive here mean that a cloudless and unreflective atmosphere with a flat reflection spectrum is plausible, since such a model would be consistent with my results to within $3\sigma$. The conclusions I draw here are only tentative, and should not be taken as a confirmation of any phenomena in the atmosphere of CoRoT-1 b.

\section{Conclusion} \label{sec:conclusion}
In this work, I searched and analyzed white-light and chromatic CoRoT data, as well as TESS data, in an attempt to find the phase variations and secondary eclipse of CoRoT-1 b. I detected the secondary eclipse in each of the CoRoT lightcurves to better than $2.4\sigma$, while in the case of the phase variations I can only confidently conclude that I have detected the phase variation in the red channel, and marginally in the blue channel, while in the white and green channels, I probably have also detected the phase variation of the planet, but that I cannot determine the true value of the phase amplitude due to residual systematic effects. In the case of the TESS data, I am unable to detect the secondary eclipse, while my choice of processing method for the TESS lightcurve leaves me unable to study the presence of any possible phase variations. I find tentative evidence that the geometric albedo of CoRoT-1 b appears to be elevated and chromatically dependent, with the albedo being higher in the green and blue CoRoT data than in the white CoRoT data, red CoRoT data, and the TESS data, suggesting that the atmosphere of CoRoT-1 b may contain a reflective cloud deck and/or exhibits Rayleigh scattering. Given the marginal significance of my fitted values, I encourage future study of the CoRoT-1 system to confirm or refute these tentative results.

\begin{acknowledgments}
This research has made use of the NASA Exoplanet Archive, which is operated by the California Institute of Technology, under contract with the National Aeronautics and Space Administration under the Exoplanet Exploration Program. This research has also made use of Lightkurve, a Python package for Kepler and TESS data analysis (\cite{2018ascl.soft12013L}).

I would also like to thank Hippke et. al. for the development of the wotan Python package, and D.J. Dorado-Daza for the ACROM set of Python scripts for processing CoRoT data. Many of the custom Python analysis scripts used in this research were derived from these two packages, and this research would not have been possible without each of them.
\end{acknowledgments}
\facilities{Exoplanet Archive, CoRoT}

\software{astropy \citep{2013A&A...558A..33A,2018AJ....156..123A}, wotan (\citet{Hippke_2019}), ACROM \citep{DoradoDaza_2018}}

\begin{figure*}
\label{fig:ecl}
\centering
\begin{subfigure}
  \centering
  \includegraphics[width=0.4\linewidth]{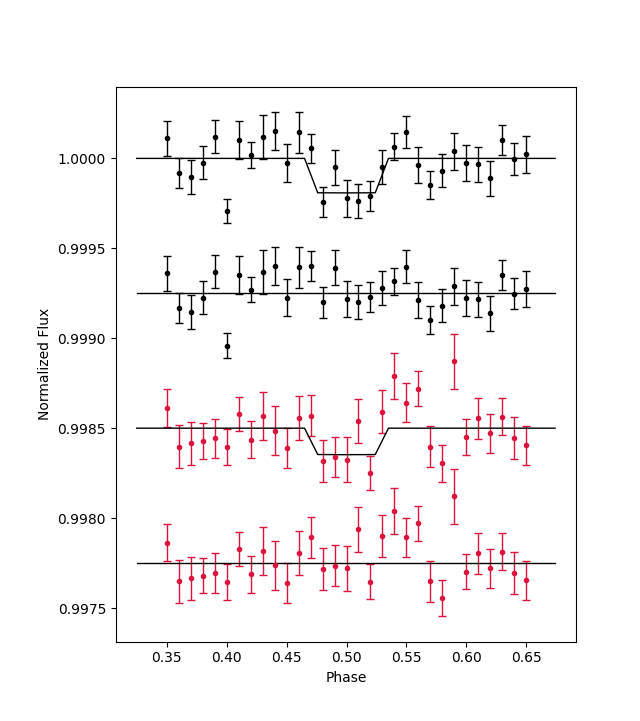}
  \label{fig:sub1}
\end{subfigure}%
\begin{subfigure}
  \centering
  \includegraphics[width=0.4\linewidth]{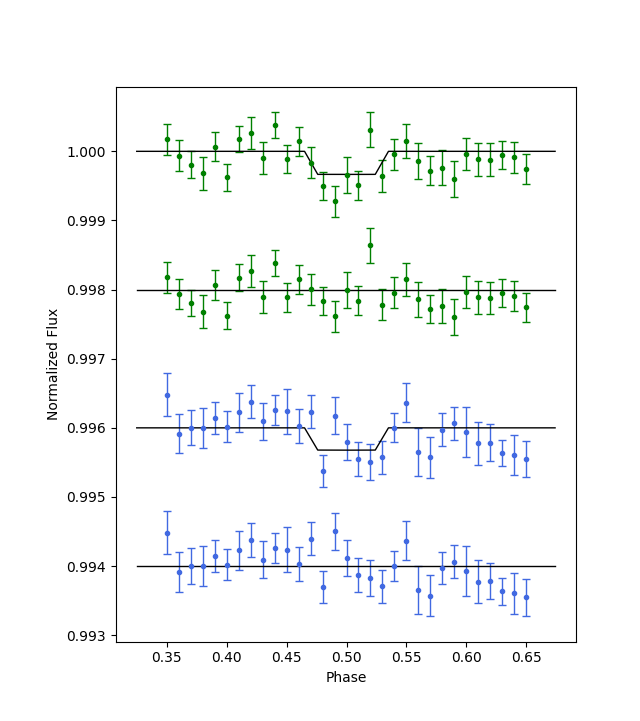}
  \label{fig:sub2}
\end{subfigure}
\caption{Best fit secondary eclipse model and residuals for the CoRoT white-light and color channel lightcurves.}
\end{figure*}

\bibliography{sample631}{}
\bibliographystyle{aasjournal}

\end{document}